\documentclass[aps,pre,twocolumn,groupedaddress]{revtex4}
\usepackage{graphicx}
\begin{document}

\title[Nonlinear dynamics of electromagnetic pulses]
{Nonlinear dynamics of electromagnetic pulses in cold relativistic 
plasmas}

\author{A. Bonatto}
\affiliation{Instituto de F\'{\i}sica, Universidade Federal do Rio Grande do
Sul, Caixa Postal 15051, 91501-970 Porto Alegre, Rio Grande do Sul,
Brasil.}
\author{R. Pakter}
\affiliation{Instituto de F\'{\i}sica, Universidade Federal do Rio Grande do
Sul, Caixa Postal 15051, 91501-970 Porto Alegre, Rio Grande do Sul,
Brasil.}
\author{F.B. Rizzato}
\email[rizzato]{@if.ufrgs.br}
\affiliation{Instituto de F\'{\i}sica, Universidade Federal do Rio Grande do
Sul, Caixa Postal 15051, 91501-970 Porto Alegre, Rio Grande do Sul,
Brasil.}

\begin{abstract}
In the present analysis we study the self consistent propagation of
nonlinear electromagnetic pulses in a one dimensional relativistic
electron-ion plasma, from the perspective of nonlinear dynamics. We show 
how a series of Hamiltonian bifurcations give rise to the electric
fields which are of relevance in the subject of particle acceleration. 
Connections between these bifurcated solutions and results of earlier 
analysis are made. 
\end{abstract}

\maketitle

\section{Introduction}

Propagation of intense electromagnetic pulses in plasmas is a subject of 
current interest in a variety of areas that make use of the available modern laser
technologies, among which we include particle and photon acceleration, nonlinear
optics, laser fusion, and others
\cite{taj79,shu86,men01,bin03,poo02,tsin99}.  Intense electromagnetic pulses
displace plasma electrons and creates a resulting ambipolar electric field
with the associated density fields. Under appropriate conditions all fields
act coherently and the pulse keeps it shape. Studies on pulse localization
have been performed in a variety of forms to unravel the corresponding
numerical and analytical properties. Kozlov et al. \cite{koz79} investigate
numerically propagation of coupled electromagnetic and electrostatic modes
in cold relativistic electron-ion plasmas to conclude that small and large
amplitude localized solutions can be present. Mofiz \& de Angelis
\cite{mof85} apply analytical approximations to the same model and suggest
where and how localized solutions can be obtained. Ensuing, more recent
works provide even deeper understanding as various features are
investigated, like influence of ion motion in slow, ion accelerating solitons
\cite{far01},  existence of moving solitons \cite{poo02}, existence of
trails lagging isolated pulses \cite{kue93,sud97} and others. Some key
points however remain not quite understood, like the way small amplitude
localized solutions are destroyed; when isolated pulses are actually free of
smaller amplitude trails; and more specific properties of the spectrum of
stronger amplitude solutions, to mention some. Those are issues of relevance
if one wishes to establish the existence range and stability properties of
the localized modes.

In the present paper we shall turn our attention to small amplitude
solitons propagating in underdense rarified plasmas, since these kind of
solitons may be of relevance for wakefield schemes. In doing so we shall follow an 
alternative strategy, other than the
direct integration of the governing equations which has been the standard approach
so far. We intend to examine the problem with techniques of nonlinear dynamics
\cite{riz97}. A canonical representation shall be constructed in
association with several tools of nonlinear dynamics like Poincar\'e maps
and stability matrices.  This strategy naturally provides a clear way to
investigate the system since we intend to establish connection between the
pulses of radiation and fixed points of the corresponding nonlinear
dynamical system \cite{lili92}. Several facts are known and
we state some which are of direct relevance for our analysis: small amplitude
solitons are created as the wave system becomes modulationally unstable at an
upper limit of the carrier frequency and cease to exist beyond a lower limit of
this carrier frequency. Not much is known on how solitons are destroyed at the
lower boundary and we examine this point to show that a series of nonlinear
resonances and bifurcations are responsible for process. A related relevant
problem is when isolated pulses are actually free of smaller amplitude trails and
this has to do with the existence of wakefields following the leading wave front 
which is of relevance for particle acceleration, for instance. Those are basic 
issues if one wishes to operate the wave system under conditions suited for 
particle acceleration, and our purpose with the present paper is to contribute 
towards the analysis of these aspects. 

\section{The model}

We follow previous works and model our system as consisting of two cold
relativistic fluids: one electronic, the other ionic. Electromagnetic
radiation propagates along the $z$ axis of our coordinate system and we
represent the relevant fields in the dimensionless forms $e {\bf
A}(z,t)/m_e c^2 \rightarrow {\bf A}(z,t)$ for the laser vector
potential, and $e \phi(z,t)/m_e c^2 \rightarrow \phi(z,t)$ for the
electric potential. $-e$ is the electron charge, $m_e$ its 
mass, and $c$ is the speed of light; $m_i$ will denote ionic mass when
appropriate. In addition, we suppose stationary modulations of a circularly
polarized carrier wave for the vector potential in the form
${\bf A}(z,t) = \psi(\tilde \xi) [{\bf{\hat x}} \sin (kz - \omega
t)+{\bf{\hat y}}
\cos (kz - \omega t)]$ with $\tilde \xi \equiv z - V t$,
whereupon introducing the expression for the vector potential into the
governing Maxwell's equation one readily obtains $V = c^2 k / \omega$. $V$
could be thus read as a nonlinear group velocity since we shall be working
in regimes where $\omega$ and $k$ are related by a nonlinear dispersion
relation.  Manipulation of the governing equations finally takes us to
the point where two coupled equations must be integrated - one
controlling the vector potential, and the other the electric potential
\cite{koz79,mof85}: 
\begin{eqnarray}
\psi'' = -{1 \over \eta} \> \psi + {V_0 \over p}\> \psi\left[{1
\over r_e(\phi,\psi)}+{\mu \over r_i(\phi,\psi)}\right], \label{equa1}\\
\phi'' = {V_0 \over p}\>\left[{(1+\phi) \over
r_e(\phi,\psi)}-{(1-\mu \phi) \over r_i(\phi,\psi)}\right]\label{equa2},
\end{eqnarray}
where the primes denote derivatives with respect to $\xi \equiv (\omega_e/c)
\> \tilde \xi$, $r_e(\phi,\psi) \equiv
\sqrt{(1+\phi)^2-p(1+\psi^2)}$, $r_i(\phi,\psi) \equiv \sqrt{(1-\mu
\phi)^2-p(1+\mu^2\psi^2)}$, $\eta \equiv \omega_e^2/\omega^2$, $\mu \equiv
m_e/m_i$, $V_0 \equiv V/c$, and $p \equiv 1-V_0^2$, with $\omega_e^2 \equiv
4\pi n_e e^2/m_e$ as the plasma frequency, and $n_e=n_i$ as the
equilibrium densities. We further rescale $\omega/ck \rightarrow \omega$
and $\omega_e/ck
\rightarrow \omega_e$ in $V_0$, $\eta$ and $p$, which helps to
simplify the coming investigation: $\eta$ preserves its form, 
$V_0 \rightarrow 1/\omega$, and $p \rightarrow 1-1/\omega^2$. A noticeable
feature of the system (\ref{equa1}) - (\ref{equa2}) is that it can be
written as a Hamiltonian system of a quasi-particle with two-degrees-of-freedom. 
Indeed, if one introduces the momenta $P_{\psi} \equiv \psi'$ and $P_{\phi}
\equiv - \phi'/p$, the equations for $\psi$ and $\phi$ takes the form
\begin{eqnarray}
\psi' = \partial H / \partial P_{\psi}, \> P_{\psi}' = - \partial H
/\partial \psi\label{equa3},\\
\phi' = \partial H / \partial P_{\phi}, \> P_{\phi}' = - \partial H /
\partial \phi\label{equa4},
\end{eqnarray} 
where the Hamiltonian $H$ reads
\begin{equation}
H={P_{\psi}^2 \over 2} - p \> {P_{\phi}^2 \over 2} + {1 \over 2 \eta} \>
\psi^2 + {V_0 \over p^2} \left[r_e(\phi,\psi)+{1\over
\mu} r_i(\phi,\psi)\right].
\label{equa5}
\end{equation}
$H$ is constant since it does not depend on the ``time'' variable
$\xi$. Its constant value, let us call it $E$, can be calculated as soon as
the appropriate initial conditions are specified. In our case we shall be
interested in the propagation of pulses vanishing for $|\xi|
\rightarrow \infty$, so we know that conditions
$P_{\psi} = P_{\phi} = \phi = \psi = 0$ must pertain to the
relevant dynamics, from which one concludes that $E=(V_0/p)^2\,(1+1/\mu)$.
Additional conditions arise from the presence of square roots in the
Hamiltonian; the dynamics lies within regions where simultaneously
$r_e^2,\>r_i^2 > 0$. Combining these inequalities with the boundary
conditions  
%
%
one is led to conclude that the entire dynamics must evolve
within the physical region 
\begin{eqnarray}
\phi_{min} \equiv \sqrt{p (1+\psi^2)} -1 < \phi<\cr
{1\over\mu}[1-\sqrt{p(1+\mu^2\psi^2)}] \equiv \phi_{max}\label{equa8}
\end{eqnarray}
if $p > 0$. If $p < 0$ there is no restriction,
but we shall see that only positive values of $p$ are of interest here.
We can also evaluate the linear frequencies of laser and
wakefield small fluctuations in the form
\begin{equation}
\psi'' = \Omega_{\psi}^2 \psi,\>\>\phi''= -
\Omega_{\phi}^2\phi,\label{equa9}
\end{equation}
where 
\begin{equation}
\Omega_{\psi}^2 \equiv - 1/\eta + 1/p \> (1+\mu),\>\>{\rm and}\>\>
\Omega_{\phi}^2 \equiv (1+\mu) / V_0^2\label{equa11}. 
\end{equation}
The potential $\phi$ oscillates
with a real frequency $\Omega_{\phi}$ which can be shown to convert into
$\omega_e (1+\mu)^{1/2}$ if dimensional variables are used for space and
time. As for  the vector potential, to reach high-intensity fields
from noise level radiation, instability must be present, which demands 
$\Omega_{\psi}^2 > 0$ and, consequently from relation (\ref{equa11}), 
\begin{equation}
1<\omega^2 \leq 1+\omega_e^2\>(1+\mu),\label{equa13}
\end{equation}
so $p>0$. 

\begin{figure}[]
\includegraphics[scale=0.5]{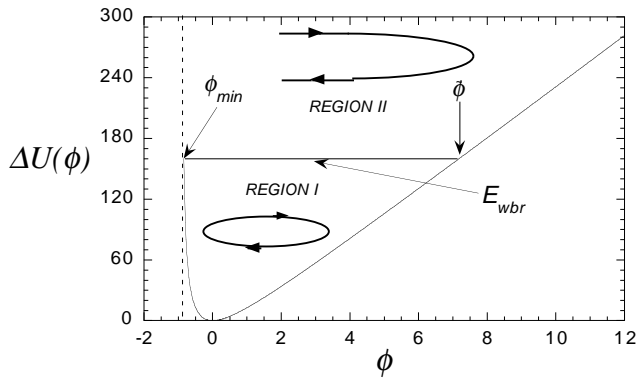}
\caption{Oscillating ($I$) and wave breaking ($II$) regions for
the electric potential at $\psi=0$. $\Delta U$ is 
defined in the text.\label{fig1.eps}}
\end{figure}

The threshold $\Omega_{\psi}^2 = 0$ can be rewritten in the
form $\omega = \omega_* \equiv \sqrt{1+\omega_e^2 (1+\mu)}$, where
$\omega_*$ is the linear dispersion relation for electromagnetic waves.
What we expect to see are small amplitude waves when $\omega$ is slightly 
smaller than $\omega_*$, with amplitudes increasing as we move farther from the
threshold. In addition to that, another feature worth of notice must be
commented. If one sits very close to the threshold, amplitude modulations
of the laser field are tremendously slow, while the oscillatory
frequency of the electric potential $\phi$ remains relatively high.
The resulting frequency disparity provides the conditions for a slow
adiabatic dynamics where, given a slowly varying $\psi$, $\phi$ always
accommodates itself close to the minimum of 
\begin{equation}
U(\phi,\psi) \equiv -{V_0 / p^2} \left[r_e(\phi,\psi)+\mu^{-1}
r_i(\phi,\psi)\right],
\label{equa13p5}
\end{equation}
the ``minus'' sign on the right hand side accounting for the negative
effective mass of $\phi$ as seen in Eq. (\ref{equa5}); note that
$\phi_{min}$ of Eq. (\ref{equa8}) refers to the smallest available
$\phi$ and not to the minimum of $U$. When
$\psi=0$, a condition to be used shortly in our Poincar\'e plots, $U$ has a minimum
at $\phi=0$ which is thus a stable point in the adiabatic regime. As one moves away from the
threshold, faster modulations and higher amplitudes may be expected to
introduce considerable amounts of nonintegrable behavior and chaos into
the system. This kind of perspective agrees well with the result of
previous works where adiabatic regions have been interpreted to be
essentially associated with small amplitude quasineutral dynamics
\cite{koz79}. One of our interests here is to precisely see how the
adiabatic dynamics is broken as one moves deeper
into nonintegrable regimes. An additional fact must be observed as one
searches for adiabatic solutions and this has to do with how close to the
minimum of $U$ on must sit so as to find these adiabatic solutions. 
The corresponding discussion parallels that on wave breaking of
relativistic eletrostatic waves. First of all note that if we do not set
$\phi$ right at the respective minimum of $U$, the electric
potential will oscillate around the minimum which will be itself displaced
due to the action of the slowly varying $\psi$. Again when $\psi =
0$, inequality (\ref{equa8}) reveals that $\phi$ must lie in the range
$\phi_{min}= \sqrt{p} -1 <0$ to $\phi_{max} = 1/\mu ( 1-\sqrt{p})>0$. Not all
these values are however actually allowed in adiabatic dynamics.
Oscillations will occur consistently only if the orbit is free to wander to
the right and left hand sides of the minimum $\phi=0$ and this can only
happen when the oscillating orbit is entirely trapped within the attracting
well of $U$. $U<0$, and a quick calculation shows that 
\begin{eqnarray}
U(\phi_{min})^2 - U(\phi_{max})^2 = 2 \sqrt{p} (1-\mu^2) \times \cr
(1-\sqrt{p})V_0^2\mu^{-2}  p^{-7/2}>0, 
\label{eq13p6} 
\end{eqnarray}
so $U(\phi_{max}) > U(\phi_{min})$, which sets a limit to cyclic orbits:
$\phi$ must be such that the corresponding potential will never be above the
level $U(\phi_{min})$. To illustrate all these comments, the reader is
referred to Fig. \ref{fig1.eps} where the potential 
$\Delta U \equiv U(\phi,\psi=0)-U(\phi=0,\psi=0)$ is represented: orbits of 
region $I$, $\phi_{min} <\phi < \tilde \phi$, will oscillate back and forth, 
but orbits in region $II$ eventually reach $\phi_{min}$ where $r_e \rightarrow 0$. 

Since it can be shown that the electronic density depends on $r_e$ in the
form $n_e
\sim r_e^{-1}$ \cite{koz79,mof85}, break down of the theory indicates wave
breaking on electrons. Also shown in the figure is the wave breaking energy 
\begin{widetext}
\begin{equation}
E_{wbr} \equiv \Delta U(\phi_{min}) = {V_o^2 \over p^2}\left[1+{1 \over \mu}-{1 \over \mu V_o} 
\sqrt{(1-\mu \phi_{min})^2 -p}\right] \approx {\omega^3 \over \omega_e^3} \> {\rm if} \> \mu,p \ll 1,
\label{ewbr}
\end{equation}
\end{widetext}
separating regions $I$ and $II$. Our conclusion is that even with extremely
slow modulations, oscillations of $\phi$ must be limited so as to satisfy the
conditions discussed above. Not only that, but the very same figure suggests
how nonintegrability affects localization of our solutions: as one moves
away from adiabaticity and into chaotic regimes, trajectories initially
trapped by $U$ may be expected to chaotically diffuse towards upper levels of this
effective potential, escaping from the trapping region, approaching $E_{wbr}$ 
and eventually hitting the boundary at $\phi_{min}$ or, in general, 
attaining $r_e=0$ for $\psi \neq 0$. If this is so, we have an explanation 
on how small amplitude solitons are destroyed, one of the issues of interest 
in the subject \cite{poo02}. We now look at the problem with help of
methods of nonlinear dynamics.

\section{Analysis with nonlinear dynamics}

\begin{figure}[]
\includegraphics[scale=0.5]{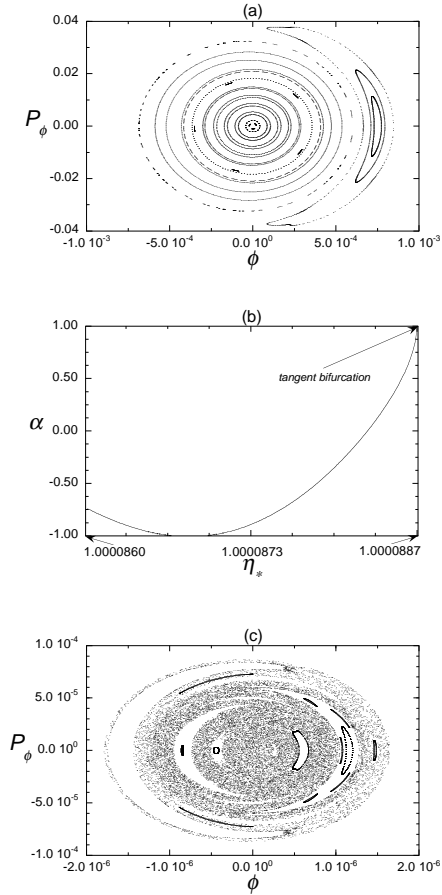}
\caption{\label{fig2.eps} (a) Phase plot
near the modulational instability threshold, with $\eta=1.00001 \eta_*$; 
(b) stability index versus $\eta$; (c) phase plot after the inverse tangency 
seen in panel (b), with $\eta = 1.0001 \eta_*$. $\epsilon=10^{-11}$.}
\end{figure}

We introduce our Hamiltonian phase space in the form of a Ponicar\'e surface
of section mapping where the pair of variables $(\phi,P_{\phi})$ is recorded each
time the plane $\psi=0$ is punctured with $P_{\psi} < 0$. Once
we have defined the map this way, we can also investigate the existence and
stability of periodic solutions of our coupled set of equations with the aid
of a Newton-Raphson algorithm. The Newton-Raphson method locates periodic
orbits and evaluates the corresponding stability index $\alpha$ which
satisfies $|\alpha| < (>) 1$ for stable (unstable) trajectories
\cite{pak01}. Parameters are represented in a form already used in earlier
investigations on the subject: we first set a numerical value for $V_0$ and
then obtain $\omega=1/V_0$ which must be larger than the unity as demanded by
condition (\ref{equa13}). However, we shall keep $V_o$ close to
the unit, and thus $\omega$ slightly larger than one, so as to represent
wave modes propagating nearly at the speed of light. This is the
convenient setting if one is interested in fast electron acceleration by
wakefields. After $V_o$ is established, the electron plasma frequency is
calculated as
$\omega_e^2 = \eta \omega^2$,
$\eta$ satisfying condition (\ref{equa13}) again. 
%
%
We note that
$\eta=\omega_e^2/\omega^2=V_0^2\omega_e^2=V_0^2\omega_{e,nonscaled}^2/c^2k^2$, so
holding $V_0$ constant while increasing $\eta$, is entirely equivalent to the more
usual practice of holding $V_0$ and the original $\omega_e$, while decreasing
$k$ and the original $\omega$. In all cases analyzed here we take $\mu = 0.0005$
as in Kozlov et al
\cite{koz79}. In addition to that, we shall take $V_o = 0.99$ to represent
the  high speed conditions of wakefield schemes. Now a crucial step is
this: since {\it isolated} pulses cannot be seen in {\it periodic} plots we alter
slightly the energy $E$ to
$E=V_0/p^2(1+1/\,\mu)\,(1+\epsilon)$, $\epsilon
\ll 1$ so the vanishing tail $P_{\psi}=P_{\phi}=\psi=\phi=0$ is avoided.
With this maneuver we convert isolated pulses into trains of {\it quasi}-isolated
pulses, a situation amenable to the use of nonlinear dynamics and the associated
periodic plots; periodicity is in fact physically meaningful if pulses result from
periodic self-modulations of initially uniform modes \cite{jos03}. In all
cases we make sure that as
$\epsilon \rightarrow 0$ the trains go into individual packets and
convergence is attained. 
%
%
The instability threshold for the vector
potential is obtained in the form $\eta_* = p/(1+\mu)=0.0198$ so 
$\omega_p \ll \omega$ as it must be in the underdense plasmas. To investigate
the adiabatic regime of the relevant nonlinear dynamics we examine phase portraits
for $\eta$ slightly larger than $\eta_*$. In panel (a) of Fig.
\ref{fig2.eps} we set $\eta=1.00001\,\eta_*$.

such a relatively small departure
from marginal stability, modulations are slow with $|\Omega_{\phi}| \gg
|\Omega_{\psi}|$, adiabatic approximations are thus fully operative and what
we see in phase space is just a set of concentric KAM surfaces rendering the
system nearly integrable. The central fixed point corresponds to 
an isolated periodic orbit since it represent a phase locked solution
that return periodically to $\psi =0$ $\phi \rightarrow 0$, and the
surrounding curves depict regimes of quasiperiodic, nonvanishing
fluctuations of $\phi$. Resonant islands are already present but still do not affect 
the central region of the phase plot where the solitary solution resides. In general 
we have observed that increasingly large resonance islands are present away 
from the central region. When $\eta$ grows the behavior of the central
fixed point can be observed in terms of its stability index represented
in Fig. \ref{fig2.eps}(b). The index oscillates
within the stable range initially, which marks the existence of a central 
elliptic point near the origin. The stability index however finally reaches
$\alpha=+1$ as indicated in the figure, beyond which point no orbit is found. 
This indicates a tangent bifurcation with a neighbouring orbit which terminates the existence 
of the central point \cite{riz02}. Immediately after tangency, the phase
plot at $\psi=0$ is still constricted to small values of $\phi$ as seen in
Fig.
\ref{fig2.eps}(c) where $\eta = 1.0001 \eta_*$. Larger values of
$\eta$ cause diffusion towards upper levels of $U(\phi)$ and we can see 
that in Fig. (3), where we investigate the behaviour of the energy 
\begin{equation}
E_{\phi} \equiv pP_{\phi}^2/2+\Delta U\label{energia}\label{enefi}
\end{equation} 
corresponding to the electrostatic field $\phi$. Instead of working directly 
with the form (\ref{enefi}) we represent diffusion in terms of compact variables 
\begin{eqnarray}
e_{\phi} \equiv {\chi_e \, E_{\phi} \over \chi_e +E_{\phi}},\label{ecomfi}\\
\Phi \equiv {\chi_{\phi} \, \phi \over \chi_{\phi} + |\phi|},\label{ficom}
\end{eqnarray}
where $\chi_e$ and $\chi_\phi$ represent the scale above which the corresponding 
variables are compactified. 

\begin{figure}[]
\includegraphics[scale=0.3]{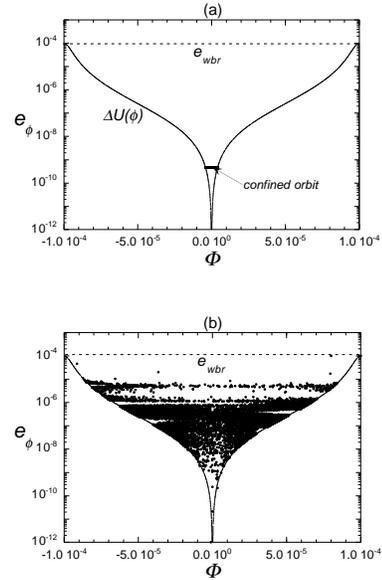}
\caption{Dynamics as represented in the $e_\phi$ versus $\Phi$ space: 
(a) $\eta = 1.00001 \eta_\ast$; (b) $\eta = 1.00021\eta_\ast$. 
$e_{wbr}\equiv \chi_e E_{wbr}/(\chi_e + E_{wbr})$.\label{fig3.eps}}
\end{figure}

\begin{figure}[]
\includegraphics[scale=0.5]{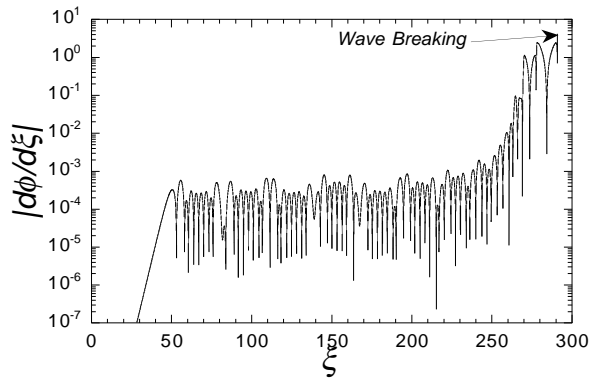}
\caption{``Time'' series for the electric field $\left|d\phi/d\xi\right|$ 
for $\eta = 1.0004 \eta_*$.\label{fig4.eps}}
\end{figure}

This kind of choice allows us to represent 
in the same plot the very extensive variations of energy and electric potential, 
without deforming these quantities when they are small, near their initial conditions. 
We found it convenient to use 
$\chi_e=\chi_\phi=0.0001$ to discuss diffusion. In Fig. \ref{fig3.eps}(a) we take 
$\eta=1.00001\eta_*$ so we are in the regular regime; as expected, no 
diffusion is observed and the quasi-particle stays near its initial condition 
$P_\phi=0$, $\phi = 10^{-8}$. For $\eta = 1.00021 \eta_*$ as in panel (b), the central 
fixed point no longer exist. In addition to that, KAM surfaces no longer isolate 
the central region of the phase plot and diffusion is observed. The quasi-particle 
moves toward $E_{wbr}$ and eventually arrives at this critical energy producing wave breaking 
on electrons. At this point the simulation stops with the electron density diverging 
to infinity. Diffusion is initially slow and becomes 
faster as energy increases. One sees voids in the diffusion plots which 
correspond to resonant islands in the phase space, so as diffusion proceeds 
the quasi-particle escalates along the contours of the resonances that become 
progressively larger as already mentioned - this is why the process is initially 
slow, becoming faster in the final stages. For larger 
values of $\eta$ no resonance is present and the quasi-particle moves quickly toward 
$E_{wbr}$. In case of panel (b) one can still see various pulses before wave 
breaking, but when $\eta$ is so large that resonances are no longer present, 
wave breaking can be instantaneous. We finally note the following relevant fact. 
For $V_o \rightarrow 1$, it is known that the amplitude of the electromagnetic 
pulses are small \cite{poo02}. But as one goes beyond the adiabatic regime, 
our discussion on diffusion allows to conclude that even small initial pulses 
eventually reach very high values, which provides the condition for formation 
of strong electric fields with the corresponding implications on particle 
acceleration. We illustrate the feature with a final figure, Fig. \ref{fig4.eps}, 
where, in a diffusive regime with $\eta = 1.0004 \eta_*$, the electric field 
$-\phi' = -pP_{\phi}$ is shown to evolve from small values near initial conditions 
to the limiting wave breaking value which agrees with the calculated value 
$|\phi'| \sim \sqrt{2\,\omega/\omega_e} \sim 3.5$.  

We read all these features as it
follows. For small enough
$\eta$'s there are locked solutions representing isolated pulses coexisting with
surrounding quasiperiodic solutions where $\phi$ does not quite vanish when 
$\psi$ does. As $\eta$ increases past the inverse tangent bifurcation but prior to full
destruction of isolating KAM surfaces, one reaches a regime of periodical
returns to $\psi=0$, although in the presence of a slightly chaotic $\phi$ motion. Those
cases where $\psi=0$ but $\phi \neq0$, correspond to
quasineutral $\psi$ pulses accompanied by trails of $\phi$ activity as
described in Refs. \cite{kue93} and \cite{sud97}. We see that trails can be
regular or chaotic. Finally, for large enough $\eta$'s, KAM surfaces no
longer arrest diffusion and wave breaking does occur as $r_e \rightarrow 0$,
as we have checked. At this point adiabatic motion is lost and this
is likely to correspond to that point where small amplitude solitary
solutions are entirely destroyed as commented in Refs. \cite{poo02} and
\cite{far01}.  

%
%

\section{Final Conclusions}

To summarize, we have used tools of nonlinear dynamics to examine the
problem of wave propagation in relativistic cold plasmas, discussing 
underdense regimes appropriate to wakefield schemes. Nonlinear dynamics provides a unified
view on the problem, thus allowing to address simultaneously several relevant
questions. In this paper we have kept our interest focused on weakly nonlinear modes
where a transition from adiabatic to nonintegrable dynamics was observed. Starting
with very low amplitude regimes near the onset of modulational instability, one
has either isolated pulses or pulses coexisting with regular
$\phi$ trails. As one increases $\eta$, thus moving away from the onset, pulses
with slightly larger amplitude exist but are never fully isolated since tangent
bifurcations annihilate the central fixed point and create ubiquitous chaotic
electrostatic trails. However, electrostatic activity is still surrounded by KAM
surfaces and therefore confined to small amplitudes. Now as one pushes amplitudes a
little higher, isolating KAM surfaces are destroyed, pulses are no longer possible
at all and wave breaking does occur. There are therefore three clearly identified
regimes in the problem: (i) regular or adiabatic regimes where the dynamics is
approximately integrable, (ii) a weakly chaotic regimes where chaos is present but
chaotic diffusion is still absent due to the presence of lingering isolating KAM
surfaces, and finally (iii) diffusive chaotic regimes where isolating KAM surfaces
are absent. Thermal effects should be added All those issues are of significant
importance for plasma accelerators and shall be developed in future publications.  
\acknowledgments
We acknowledge partial support by CNPq, Brasil.
%
%

%
\end{document}